\documentclass{aa}  
\usepackage{amsmath}
\usepackage{amssymb}
\usepackage{graphicx}
\usepackage{color}
\usepackage{subfigure}
\usepackage{bm}
\usepackage{textcomp,gensymb}
\usepackage{bigints}
\usepackage[dvipsnames,svgnames,table]{xcolor}
\usepackage{url}
\usepackage{orcidlink}
\usepackage{physics}
\usepackage{chemformula}
\usepackage{adjustbox}
\usepackage{booktabs}

\usepackage{hyperref}
\hypersetup{
    colorlinks=true,
    linkcolor=Blue,
    filecolor=Blue,
    urlcolor=Blue,
    citecolor=Blue
}
\begin{document}

   \title{Structural quasi-universality in highly magnetized differentially rotating neutron stars}
    \titlerunning{Structural quasi-universality in highly magnetized differentially rotating neutron stars}

   \author{Pratul Manna \orcidlink{0009-0006-4918-2943},
          \inst{1}
          \thanks{\email{mpratul@astrouw.edu.pl}}
          \and
          Surajit Kalita \orcidlink{0000-0002-3818-6037},
          \inst{1}
          \thanks{\email{skalita@astrouw.edu.pl}}
          }
   \institute{
             Astronomical Observatory, University of Warsaw, Al. Ujazdowskie 4, Warszawa 00-478, Poland
             }
    \authorrunning {P. Manna \& S. Kalita}

   \date{Received August 25, 2026; accepted September 24, 2026}


\abstract
   {Universal relations among macroscopic properties of neutron stars provide a powerful framework to probe their internal structures while minimizing uncertainties associated with the equation of state (EoS). Although such relations have been extensively studied for uniformly rotating stars, their extension to differentially rotating and strongly magnetized configurations is relatively less explored.}
   {We aim to investigate whether it is possible to obtain a phenomenological quasi-universal relation between moment of inertia and stellar compactness for differentially rotating strongly magnetized neutron stars modeled with nuclear equation of states.}
   {We systematically investigate equilibrium configurations for a wide range of EoSs, rotation profiles, and magnetic field strengths with \textsc{xns} numerical simulation code. For sequences with fixed angular momentum, we analyze the dependence of normalized moment of inertia on compactness.}
   {We establish a generalized quasi-universal relation between the moment of inertia and compactness that remains remarkably insensitive to the underlying EoS across uniformly rotating, differentially rotating, and strongly magnetized toroidal configurations. For sequences with fixed angular momentum, the normalized moment of inertia exhibits a quasi-universal dependence on compactness, with deviations primarily due to magnetic field strength and degree of differential rotation. We derive analytic expressions for these dependencies, enabling a unified phenomenological model applicable over a wide range of stellar configurations. As astrophysical applications, we quantify the systematic bias in magnetar luminosity estimates and the rotational kinetic energy of post-merger remnant of GW170817.}
   {These results extend quasi-universal relations beyond the standard assumptions of uniform rotation and weak magnetization, providing a robust framework for interpreting observations of highly magnetized and differentially rotating neutron stars, and enabling more reliable constraints on their astrophysical properties.}

\keywords{Stars: neutron -- Stars: magnetars -- Stars: magnetic field -- Stars: rotation -- Gravitation -- Gravitational waves}

\maketitle
%

\section{Introduction}

Neutron stars (NSs) are the most compact and strongly magnetized stellar remnants currently known, making them unique astrophysical laboratories for probing the behavior of matter under conditions of extreme densities. In the era of multimessenger astronomy~\citep{LIGOScientific:2017ync}, a deeper understanding of NS matter is crucial for advancing the fundamental physics through forthcoming gravitational wave (GW) detections and electromagnetic (EM) observations. Binary NS coalescences constitute one of the most promising source classes for the third-generation GW detectors like the Einstein Telescope and Cosmic Explorer~\citep{Evans:2021gyd}.

Core-collapse supernovae and binary mergers are among the typical formation channels through which proto-NSs are formed. The post-formation stability of these remnants is strongly governed by their rotational state~\citep{Cook1992ApJ,Cook1994ApJB,Cook1994ApJA}. In realistic environments, differentially rotating remnants can be stabilized temporarily against prompt collapse to a stellar-mass black hole~\citep{Baumgarte:1999cq}. Numerical simulations of such configurations demonstrated that differential rotation can significantly affect the macroscopic properties of NSs (e.g. maximum mass and circumferential radius) leading to noticeable deformations of the objects~\citep{Studzinska:2016ofb,Gondek-Rosinska:2016tzy,Espino:2019ebx,Szkudlarek:2019odl}. As a consequence, it can leave imprints on the observed GW signal during collapse. Therefore, studying such objects are of utmost importance in anticipation to Galactic supernovae expected to be detected by current and near-future instrumentation~\citep{szczepanczyk}.

One subclass of NSs, observationally identified as Soft Gamma Repeaters and Anomalous X-ray Pulsars~\citep{kat1998Natur.393..235K,Gavriil:2002mc}, are believed to possess strong magnetic fields. These objects, commonly referred as `magnetars', are characterized by surface dipole field strengths inferred from spin-period measurements ($P$, $\dot{P}$)~\citep{duncan1992ApJ,duncan1993ApJ,duncan1995MNRAS.275..255T,duncan1996ApJ,Kaspi:2017fwg}, while virial theorem arguments indicate that internal field strengths may reach upto $10^{18}$\,G~\citep{Chatterjee:2018prm}. Although, the current catalog of magnetars comprises of a limited number of observations~\citep{Olausen}\footnote{\url{https://www.physics.mcgill.ca/~pulsar/magnetar/main.html}}, they are expected to represent a considerable fraction of population of NSs in the future~\citep{Kaspi:2017fwg}. Therefore, apart from rotation (uniform or differential), such strong magnetic fields can impact the geometry of the objects~\citep{Wei:2017luh,Chatterjee:2018prm}. In particular, uniform rotation and poloidal field configurations induce oblate deformations, toroidal fields produce prolate geometries, and strong differential rotation can generate a polar-hollow morphology.

Differential rotation plays a crucial role in the evolution of NSs because it can efficiently amplify seed magnetic fields through magnetic winding and the magnetorotational instability (MRI) \citep{Reboul-Salze:2020mnw}. Differential rotation also redistributes angular momentum in the central regions of rapidly rotating proto-NSs and merger remnants and could account for magnetar field strengths. Full general relativistic numerical simulations showed that magnetic braking and MRI drive differentially rotating NSs toward uniform rotation on relatively short timescales~\citep{Duez:2006qe}. In hypermassive NSs formed during binary mergers, this magnetic angular momentum transport can trigger delayed collapse into a black hole surrounded by a magnetized accretion torus. More recently, magnetohydrodynamic (MHD) studies have further shown that strong internal magnetic fields can spontaneously generate differential rotation even in initially uniformly rotating magnetars~\citep{Tsokaros:2021pkh}.

Several efforts have been going on constructing equation of state (EoS) that describes the bulk properties and internal composition of NSs. However, different uncertainties persist regarding the behavior of matter at supra-nuclear densities where the stellar core density can exceed $10^{15}\rm\,g\,cm^{-3}$. Owing to limited understanding of microscopic effects such as strong interactions, phase transitions, exotic degrees of freedom, etc., theoretical predictions of NS observables often exhibit strong dependence on the underlying EoS model. To mitigate this effect, several approximate universal relations have been developed over the past decades~\citep{ravenhall,Lattimer:2000nx,Bejger:2002ty,Yagi:2013bca,Breu:2016ufb}, providing a robust framework for interpreting astrophysical observations in a comparatively EoS-insensitive manner. Most existing studies have focused on the effects of uniform rotation on such relations, with extensive investigations of quantities such as the moment of inertia, tidal deformability, quadrupole moment, and compactness. In comparison, the effects of differential rotation and magnetic fields on these properties remain comparatively less explored. Understanding whether universal relations remain valid in the simultaneous presence of differential rotation and magnetic fields is therefore still an active area of research.

Over the years, several numerical codes have been developed to construct equilibrium models of rotating and magnetized NSs. The \textsc{rns} code \citep{1989MNRAS.237..355K} was developed to study relativistic rotating stellar models, while the \textsc{lorene} code\footnote{\url{https://lorene.obspm.fr}} uses spectral methods to construct stationary, axisymmetric configurations, including those with magnetic fields. However, each code depends on specific assumptions about the magnetic field geometry and rotation law. However, constructing equilibrium models that simultaneously incorporate differential rotation and magnetic fields requires a formulation that treats both effects self-consistently within the equilibrium equations. The \textsc{xns} code \citep{Bucciantini_2011} is designed precisely for this purpose and is widely used in this field. For example, it has been employed to investigate different rotation profiles for binary NS remnants~\citep{Franceschetti:2022ypc}, connecting magnetic deformations with continuous GWs~\citep{Soldateschi:2021hfk}, developing universal relations~\citep{Soldateschi:2021cxq}, MHD modeling of NSs~\citep{DelZanna:2017umb} and many other contexts~\citep{Das:2014owa, Camelio:2019rsz, Soldateschi:2020hju, Camelio:2020mdi, 2021MNRAS.508..842K, Das:2026ffp, Pinas:2025bpq}. It constructs axisymmetric NS equilibrium configurations and supports uniform and differential rotation alongside purely toroidal, purely poloidal, and twisted-torus magnetic field configurations. These capabilities make \textsc{xns} more suitable tool for the models considered in this work, which combine differential rotation with toroidal magnetic fields. Other numerical codes, such as \textsc{Gmunu} \citep{2021MNRAS.508.2279C}, \textsc{Einstein Toolkit}\footnote{\url{https://einsteintoolkit.org}}, \textsc{CoCoNuT}~\citep{cedra_duran} should be distinguished from these equilibrium solvers, as they are primarily designed for the subsequent time-dependent dynamical evolution of relativistic magnetized fluids~\citep{Matur:2026hpz, Pinas:2025bpq, Varma:2026swh} rather than for constructing the equilibrium configurations themselves.

Universal relations for NSs in the presence of strong magnetic fields are relatively less studied than their non-magnetized counterparts. \cite{Soldateschi:2021cxq} made important progress by deriving quasi-universal scaling relations connecting the magnetic deformability of static, axisymmetric NSs with their Komar mass, circumferential radius, and magnetic field strength for both purely poloidal and purely toroidal field geometries. They showed that after a suitable normalization, the magnetic deformation is largely insensitive to EoS. Their analysis was performed for equilibrium configurations computed using the \textsc{xns} code in general relativity and was further extended to scalar-tensor theories. 

The present work addresses a complementary problem. We analyze the combined effects of differential rotation and toroidal magnetic fields on the macroscopic properties of NSs using two fundamental observables: moment of inertia ($I$) and stellar compactness ($\mathcal{C}$). Based on analysis of an extensive set of equilibrium configurations spanning various physical parameters, we derive a quasi-universal relation that accounts for both differential rotation and magnetic fields, while maintaining largely independent of underlying EoS. Specifically, for sequences of constant angular momentum ($J$), we identify a simple relation between normalized $I$ and $\mathcal{C}$ that is independent of the EoS and is parametrically determined by the magnetic field strength and the degree of differential rotation. In addition, we provide explicit functional forms for the dependence of the model parameters on these quantities; thereby extending the applicability of this unified relation to a broader class of compact objects. We use this relation as proof-of-concept applications to estimate astrophysical observables like spin-down luminosity of magnetars and post-merger rotational energy of remnants like GW170817. Within the scope and limitations of our framework, we find that, ignoring differential rotation and magnetic field effects can induce systematic uncertainties in the inferred quantities.

The article is structured as follows. In Sec.~\ref{sec:methodology}, we describe the theoretical framework and numerical tools used to construct equilibrium configurations of rotating magnetized NSs. Sec.~\ref{sec:unified_rel} provides a description of step-by-step development of the unified relation. In Sec.~\ref{sec: astro_imp}, we use the relation to draw constraints on the observable properties using publicly available data of magnetars and properties of GW170817 NS merger event. Finally, in Sec.~\ref{sec: discussion}, we discuss the findings and limitations of this work.


\section{Methodology}\label{sec:methodology}
\subsection{Theoretical Framework}

In spherical coordinates $(t,r,\theta,\phi)$, the spacetime of a static, axisymmetric compact star can be accurately described within the conformal flatness condition (CFC) approximation in quasi-isotropic coordinates~\citep{oron:2002,Shibata:2005gp,Dimmelmeier:2005zk,Ott:2006eh,Bucciantini_2011,Pili:2014npa,Pili:2017yxd,Soldateschi:2021cxq}. Under this approximation with the assumption $G=c=1$ with $G$ being the Newton gravitational constant and $c$ the speed of light, the line element takes the form
\begin{equation}
    \dd{s}^2=-\alpha^2 \dd{t}^2 + \psi^4\left[\dd{r}^2+r^2\left(\dd\theta^2+\sin^2\theta \dd\phi^2\right)\right],
\end{equation}
where $\alpha$ and $\psi$ denote the lapse function and conformal factor respectively. In general $\alpha$ and $\psi$ are functions of both space and time. However, for a static axisymmetric spacetime these functions depend only on the spatial coordinates $r$ and $\theta$, and are independent of $t$ and $\phi$. The matter content is modeled as a magnetized ideal fluid in the framework of ideal general-relativistic-MHD (GRMHD), where the EM field satisfies
\begin{equation}
    F^{\mu\nu}u_\nu = 0,
\end{equation}
with $F_{\mu\nu}$ and $u_\mu$ being the Faraday tensor and four velocity, respectively. It implies that the electric field measured in the comoving frame vanishes and the Faraday tensor is entirely determined by the magnetic field four vector. The stress-energy tensor is then given by~\citep{Bucciantini_2011}
\begin{equation}
    T^{\mu\nu}=\rho h u^\mu u^\nu + Pg^{\mu\nu}+F^\mu_\delta F^{\nu\delta}-\frac{1}{4}F^{\delta\gamma}F_{\delta\gamma}g^{\mu\nu},
\end{equation}
where $h=1+\epsilon+P/\rho$ is the specific enthalpy. Here $g_{\mu\nu}$, $\epsilon$, $\rho$, and $P$ denote the metric tensor, energy density, rest mass density, and pressure, respectively. The relations between different thermodynamic variables of dense nuclear matter are determined by the underlying EoS. Thereby stellar configurations are obtained by solving the GRMHD equations along with Einstein's field equations under axisymmetric equilibrium conditions~\citep{Bucciantini_2011}.

For rotating configurations, the integrability condition of the equilibrium equations requires the specification of a rotation law $F(\Omega)$, where $\Omega$ is the angular velocity. In this work, we adopt the Komatsu-Eriguchi-Hachisu (KEH) differential rotation profile, commonly referred to as the $J-$constant law \citep{Komatsu:1989ikr,1985A&A...146..260E}
\begin{equation}
    F(\Omega)=A^2(\Omega_c-\Omega),
\end{equation}
where $\Omega_c$ denotes the central angular velocity and $A$ characterizes the degree of differential rotation. Physically, $A$ defines the length scale over which the angular velocity varies within the star, with the limit $A\rightarrow\infty$ corresponding to uniform rotation. This profile produces a maximum $\Omega$ at the center that decreases monotonically toward the surface. The KEH law has been shown to provide a reasonable description of differentially rotating remnants formed in core-collapse supernovae~\citep{Villain:2003ey} and has been widely used in dynamical merger simulations~\citep{Baumgarte:1999cq,Lyford:2002ip}.

The magnetic field effects are modeled using the simple barotropic law~\citep{Kiuchi:2008ch,Bucciantini_2011}
\begin{equation}
    \alpha\mathcal{R}B=K_m(\alpha^2\mathcal{R}^2\rho h)^m,
    \label{mag_law}
\end{equation}
where $\mathcal{R}=\psi^2 r\sin\theta$ denotes the generalized cylindrical radius. The magnetic field strength ($B$) is determined by two control parameters: magnetization constant ($K_m$) and magnetic index ($m$). Together, these parameters determine both the strength and spatial profile of the magnetic field inside the star.
\subsection{Numerical setup}

We use the open source code \textsc{xns}\footnote{\url{https://www.arcetri.inaf.it/science/ahead/XNS/code.html}} to obtain stationary axisymmetric equilibrium configurations under the eXtended Conformal Flatness condition~\citep{Cordero-Carrion:2008grk}. We use 600 stationary, axisymmetric stellar models constructed with 10 tabulated hadronic EoSs for which the maximum non-rotating masses range between $\sim2-3\rm\,M_\odot$ (See Table~\ref{tab:eos} for details). For magnetized systems, in the barotropic law of Eq.~\eqref{mag_law}, we fix $m=1$ and vary $K_m$ such that the resulting maximum internal magnetic field strength ($B_{\rm max}$) lies in the range $\left[10^{17}-10^{18}\right]$\,G. Throughout this work, we restrict our analysis to purely toroidal magnetic field configurations, with no poloidal magnetic field component included. This field configuration induces prolate deformations of the stellar structure, as opposed to the oblate geometries associated with poloidal fields. A direct consequence of the absence of poloidal fields is that the magnetic fields are entirely confined within the stellar interior and the resulting equilibrium configurations do not support magnetic fields extending to the stellar exterior like magnetosphere or corona. Nevertheless, equilibrium configurations of mixed poloidal--toroidal fields were investigated in previous studies using \textsc{xns} code by \cite{2015MNRAS.447.2821P} and \cite{Soldateschi:2021cxq}, which represent a different class of equilibrium configurations from those considered here.

\begin{table}[htbp]
\centering
\caption{Details of EoSs used in this work. $M_{\rm max}$ denotes the maximum mass of the static configuration obtained from the \texttt{CompOSE} database\tablefootmark{a}.}
\label{tab:eos}
\setlength{\tabcolsep}{6pt}
\begin{tabular}{|c|c|c|c|}
\hline
{Count} & {EoS} & Technique & $M_{\rm max}~\rm (M_\odot)$ \\
\hline
1 & APR   & nuclear many-body  & 2.19 \\
2 & SLy4  & Skyrme mean-field        & 2.06 \\
3 & SKI3  & Skyrme mean-field        & 2.25 \\
4 & SK272 & Skyrme mean-field        & 2.24 \\
5 & SFHo  & Relatvistic mean-field   & 2.06 \\
6 & SFHx  & Relatvistic mean-field   & 2.21 \\
7 & DD2   & Relatvistic mean-field   & 2.42 \\
8 & NL3   & Relatvistic mean-field   & 2.79 \\
9 & TMA   & Relatvistic mean-field   & 2.02 \\
10 & TM1  & Relatvistic mean-field   & 2.21 \\
\hline
\end{tabular}
\tablefoottext{a}{\url{https://compose.obspm.fr}}
\end{table}

It is well established that purely poloidal and purely toroidal field configurations are generally susceptible to MHD instabilities, with the former to Tayler-type instabilities and the latter to convective and kink-mode instabilities~\citep{2009MNRAS.397..763B}. In contrast, twisted-torus configurations with mixed poloidal--toroidal components are widely regarded as among the most physically plausible candidates for stable equilibrium in NSs, as they allow mutual stabilization between the two field components. Such twisted-torus configurations, including those in which the magnetic field extends continuously into the stellar exterior and satisfies appropriate matching conditions at the surface, have been self-consistently constructed using the \textsc{xns} code~\citep{2015MNRAS.447.3278B,2015MNRAS.447.2821P}. The present work should therefore be viewed as complementary to these studies, focusing specifically on the properties of toroidal field component.


Nevertheless, this limitation does not significantly affect our results, as we focus on global equilibrium properties that are predominantly determined by the quantities in stellar interior, such as the central density and central magnetic field strength. Numerical simulations of differentially rotating NSs have shown that the toroidal magnetic field component can dominate over the poloidal component in the central regions by more than two orders of magnitude on dynamical timescales~\citep{Wickramasinghe_2013,2021MNRAS.508..842K}. Therefore, purely toroidal configurations provide a reasonable approximation for studying the influence of strong interior magnetic fields on bulk stellar properties (see for example~\citealt{Frieben:2012dz}).

\section{The unified relation}\label{sec:unified_rel}

In this section, equilibrium sequences of constant $J$ have been constructed to study the relation between the scaled moment of inertia $\tilde{I}=I/M^3$ and compactness $\mathcal{C}=M/R$ of NSs, where $M$ is the mass of the NS and $R$ is its equatorial radius. For each configuration, $J$ is held fixed at a constant value $J_\mathrm{const} = 0.1968$ by iteratively adjusting the central angular velocity $\Omega_c$. Following \cite{Taylor:2019hle}, we begin with an initial guess for $\Omega_c$ and compute the resulting angular momentum $J_0$ using the \textsc{xns} code. As $J\propto \Omega_c$, we rescale $\Omega_c$ by the factor $J_\mathrm{const}/J_0$ and repeat the procedure until the target $J_\mathrm{const}$ is achieved. Furthermore, we consider only those configurations which have moderate spin frequency $j = J/M^2 \lesssim 0.35$.\footnote{The choices of $J_{\rm const}$ and $j$ are relevant for NS configurations residing on the stable branch of $J-$constant sequences and applicable for scenarios like differentially rotating proto-NS and fast milisecond pulsars.} In particular, it was shown that the accuracy of slow-rotation fits reduces as $j$ increases for uniformly rotating NSs \citep{Breu:2016ufb}. Finally, for each EoS, we focus only on systems residing on the stable branch of $J-$constant sequences following the turning point criterion~\citep{Friedman:1988er}.

In this work, we first examine the applicability of the slow-rotation fitting formula proposed by \cite{Breu:2016ufb} and test its validity for uniformly rotating NSs across the EoSs considered here. We then quantify the effects of differential rotation on these sequences while keeping $J$ fixed, and accordingly modify the $\tilde{I}-\mathcal{C}$ relation. Finally, we incorporate strong toroidal magnetic fields into the equilibrium models and present a unified relation that accounts for rotation, differential rotation, and magnetization.

\subsection{Rigid rotation}
Before investigating the effects of differential rotation and magnetic fields, it is necessary to first validate the established universal relations for uniformly rotating NSs. As proposed by \cite{Breu:2016ufb}, the universal relation after expressing $\tilde{I}$ as an inverse power series in $\mathcal{C}$ is given below:
\begin{equation}
    \tilde{I} = \sum_{i=1}^4a_i\mathcal{C}^{-i}.
    \label{Eq: rigid_fit}
\end{equation}
The choice of an inverse-power expansion is motivated by the leading-order scaling $I/M^3 \sim \mathcal{C}^{-2}$. The coefficients $a_1$, $a_2$, $a_3$, and $a_4$ were originally determined by fitting equilibrium sequences characterized by fixed $j$ (see \citealt{Breu:2016ufb} for more details).

\begin{figure}[htbp]
    \centering
    \includegraphics[width=\columnwidth]{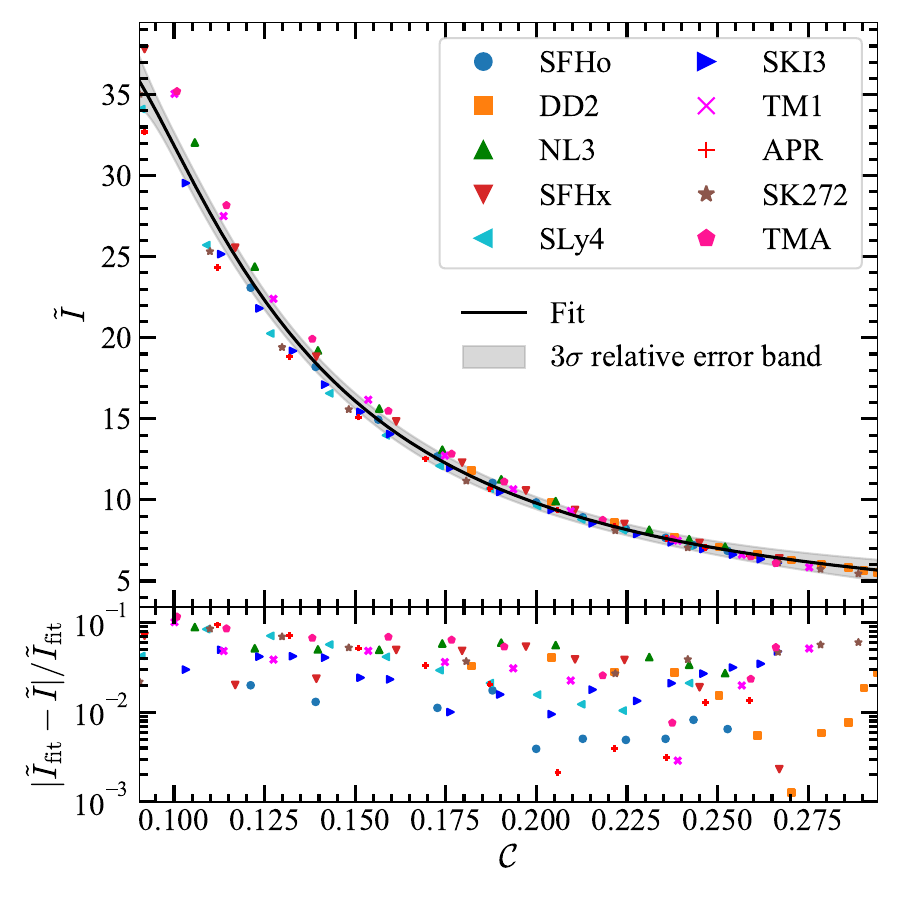}
    \caption{\textit{Top}: Normalized moment of inertia $\tilde{I}(=I/M^3)$ plotted as a function of compactness $\mathcal{C}(=M/R)$ for 10 EoSs listed in Table~\ref{tab:eos}. The black solid line represents the universal relation and the shaded grey region denotes $3\sigma$ relative error with respect to the fitted curve. \textit{Bottom}: Relative error plotted for all the configurations considered here.}
    \label{Fig: rigid_rot_uni_rel}
\end{figure}

In this work, we adopt the same fitting prescription and determine the corresponding coefficients using the set of EoSs considered here. Fig.~\ref{Fig: rigid_rot_uni_rel} shows the dependence of $\tilde{I}$ on $\mathcal{C}$ for uniformly rotating NSs constructed with different EoSs, together with the fit obtained from Eq.~\eqref{Eq: rigid_fit}. It is evident that all models nearly lie within the $3\sigma$ relative-error band of the fitted relation, irrespective of the underlying EoS; thereby confirming the robustness of the universal behavior. The resulting fitting coefficients are listed in the first row of Table~\ref{Tab: Fitted coeffictient B=0}. Note that the fitted coefficients differ slightly from those reported by \cite{Breu:2016ufb}, primarily due to the different set of EoSs and selected range of $j$ considered in the present analysis. Nevertheless, the relative deviations from the universal relation are small (typically below $\sim10\%$), as illustrated in the bottom panel of Fig.~\ref{Fig: rigid_rot_uni_rel}.  

\begin{table}[htbp]
    \centering
    \caption{Table of fitted coefficients of Eq.~\eqref{Eq: rigid_fit} for different degrees of differential rotation.}
    \begin{tabular}{|l|c|c|c|c|c|}
        \hline
         & $a_1$ & $a_2$ & $a_3$ & $a_4$ & $\chi^2_{\rm red}$\\
        \hline
       $A\to\infty$ & $2.1371$ & $-0.4486$ & $0.1099$ & $-0.0055$ & $0.0019$\\
        \hline
       $A=10$ & $1.2859$ & $-0.2972$ & $0.0855$ & $-0.0047$ & $0.0029$\\
        \hline
       $A=8$ & $0.6669$ & $-0.0391$ & $0.0389$ & $-0.0022$ & $0.0048$\\
        \hline
       $A=6$ & $0.1648$ & $0.1437$ & $0.0020$ & $-0.0002$ & $0.0089$\\
        \hline
    \end{tabular}
    \label{Tab: Fitted coeffictient B=0}
\end{table}

\newpage
\subsection{Differential rotation}
\begin{figure}[htbp]
     \subfigure[Soft EoS (SLy4).]{%
       \includegraphics[width=0.98\columnwidth]{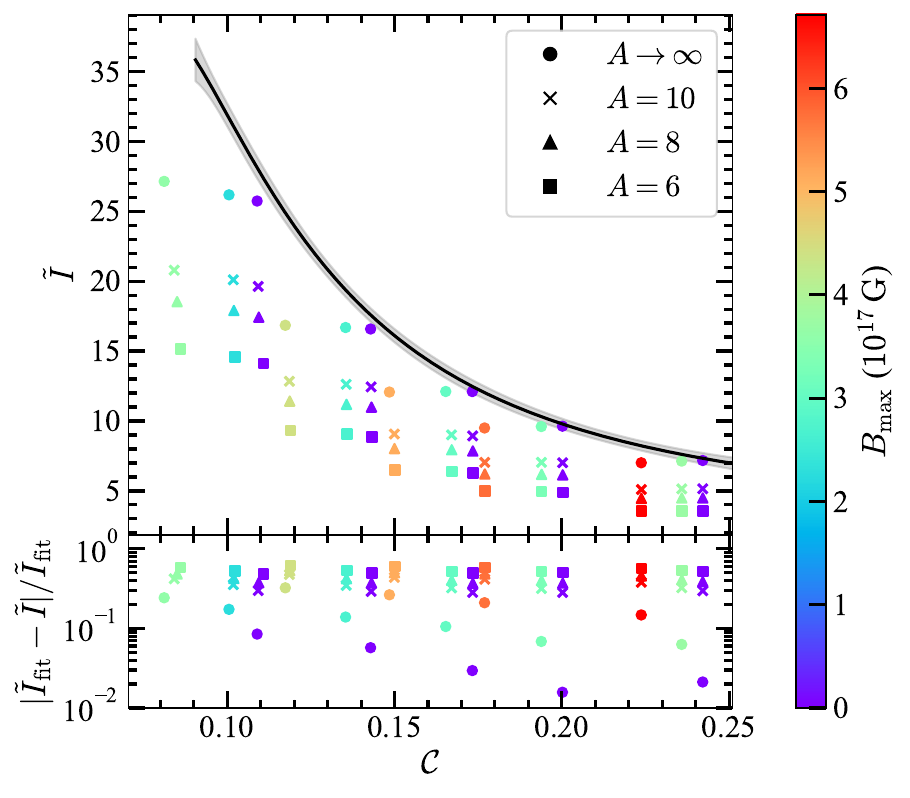}
    }\\
     \subfigure[Stiff EoS (NL3).]{%
       \includegraphics[width=0.98\columnwidth]{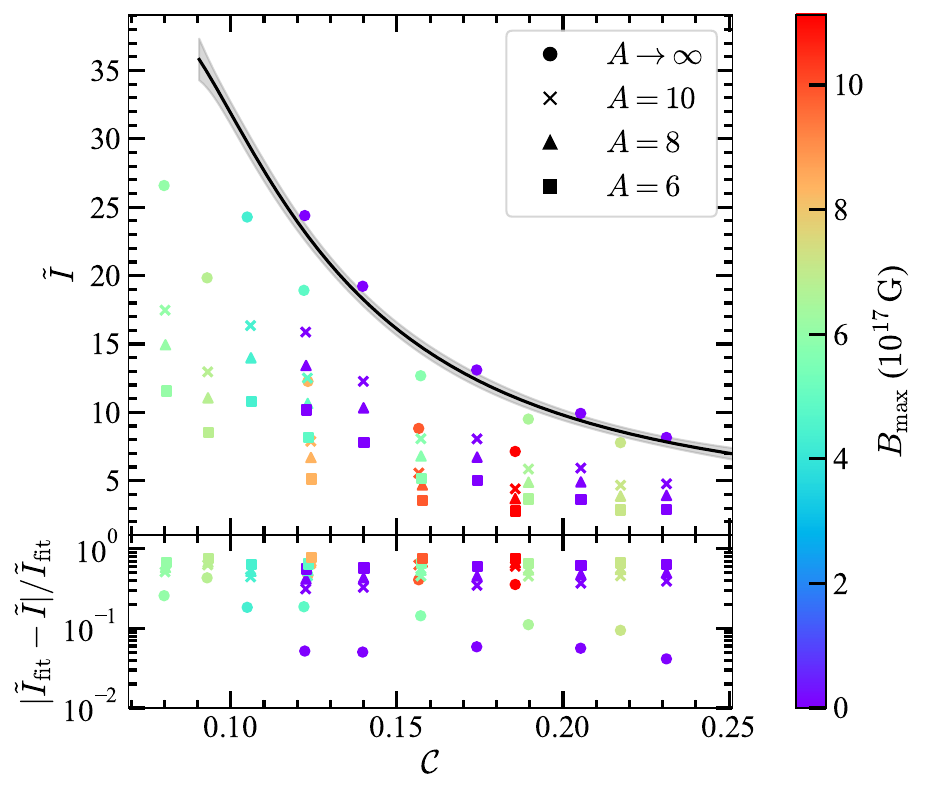}
    }
    \caption{$\tilde{I}$ as a function of $\mathcal{C}$ for differentially rotating NSs in the presence of toroidal magnetic fields for 2 EoSs. The color bar shows maximum interior magnetic field ($B_{\rm max}$) of the NS. The relative deviations from slow rotation fit are shown in the bottom panel of each plot.}
    \label{Fig: mag_field_effects}
\end{figure}

We now investigate the impact of differential rotation on the equilibrium configurations. We consider three values of the differential rotation parameter $A =6, 8, 10$.\footnote{We parametrize the equilibrium sequences using $A$ as a direct input to the \textsc{xns} code, since it uniquely determines $J$ distribution without requiring any iterative adjustment. Fixing $A$ therefore ensures a consistently defined rotation law throughout the sequence and facilitates direct comparisons among different EoSs. Although the dimensionless ratio $A/r_\mathrm{e}$ (where $r_\mathrm{e}$ denotes the equatorial radius) provides a more physically intuitive measure of the degree of differential rotation, $r_\mathrm{e}$ itself is an output of the equilibrium solver. Hence, $A/r_\mathrm{e}$ cannot be used as a control parameter for constructing equilibrium sequences.} Among the 10 EoSs considered, we present only two representative cases in Fig.~\ref{Fig: mag_field_effects} where panel~(a) corresponds to the soft SLy4 EoS and panel~(b) corresponds to the stiff NL3 EoS. The configurations are constructed at constant $J$ and in this case, the relative deviations can reach as high as $\sim 18\%$.

As shown in the figure, decreasing $A$ (i.e. increase in degree of differential rotation) leads to gradually larger deviation from the uniform rotation $\tilde{I}-\mathcal{C}$ relation. More specifically, for a fixed $\mathcal{C}$, differentially rotating models systematically exhibit smaller values of $\tilde{I}$ over the entire range of central densities considered (see Sec.~\ref{sec: discussion} for more details). Similar trends are observed for all EoSs included in this study. The best-fit coefficients $a_1-a_4$ are listed in Table~\ref{Tab: Fitted coeffictient B=0} for each value of $A$.
It is evident that in order to accommodate the effects of differential rotation, it is essential to upgrade the polynomial coefficients of Eq.~\eqref{Eq: rigid_fit} as functions of $A$, i.e. $a_i=a_i(A)$. To determine the functional dependence, we model each coefficient $a_i$ using a third-order polynomial expansion in inverse powers of $A$, i.e.
\begin{equation}
    a_i(A)= \sum_{i=0}^3 n_i A^{-i}. 
    \label{a_i_mapping}
\end{equation}
Note that $n_1$, $n_2$, and $n_3$ are fitting parameters which accounts for the characteristic length scale associated with the differential rotation profile. Fitting Eq.~(\ref{a_i_mapping}) to the values listed in Table~\ref{Tab: Fitted coeffictient B=0} yields the following expressions\footnote{We verify that for $A \gtrsim 20$, the influence of differential rotation becomes negligible, and the corresponding $\tilde{I}-\mathcal{C}$ relation converges to that obtained for uniformly rotating configurations.}
\begin{equation}\label{Eq: a coeff}
    \begin{aligned}
        a_1 &= 2.0895 + \frac{12.9651}{A} - \frac{316.8588}{A^2} + \frac{1014.7281}{A^3}, \\
        a_2 &= -0.4345 - \frac{3.3118}{A} + \frac{69.6192}{A^2} - \frac{171.1361}{A^3}, \\
        a_3 &= 0.1075 + \frac{0.5409}{A} - \frac{10.9518}{A^2} + \frac{23.0102}{A^3}, \\
        a_4 &= -0.0053 - \frac{0.0225}{A} + \frac{0.3967}{A^2} - \frac{0.4320}{A^3}.
    \end{aligned}
\end{equation}
In each case, the resulting fit yields a reduced $\chi^2$ value below 0.02, indicating excellent agreement of the fitting function with the numerical data.

\subsection{Magnetic fields}
\label{mag_fields}
\begin{figure}[htbp]
     \subfigure[$\tilde{I}(\mathcal{C},B_\mathrm{max})$ as a function of $\mathcal{C}$.]{%
       \includegraphics[width=0.98\columnwidth]{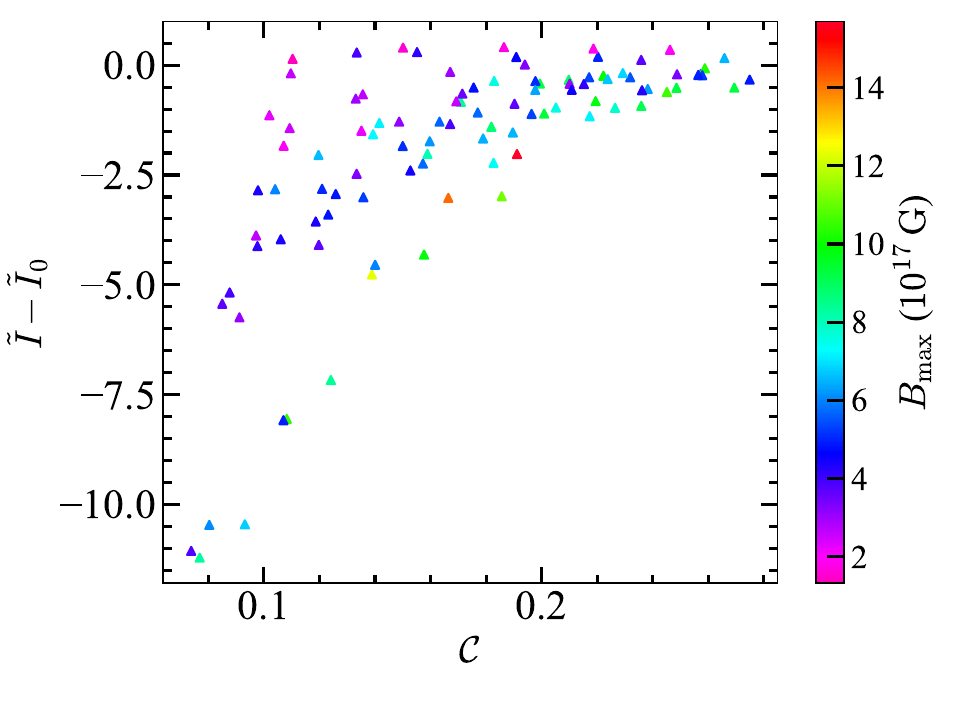}
    }\\
     \subfigure[$\tilde{I}(\mathcal{C},B_\mathrm{max})$ as a function of $\mathcal{C}/\sqrt{\epsilon}$.]{%
       \includegraphics[width=0.98\columnwidth]{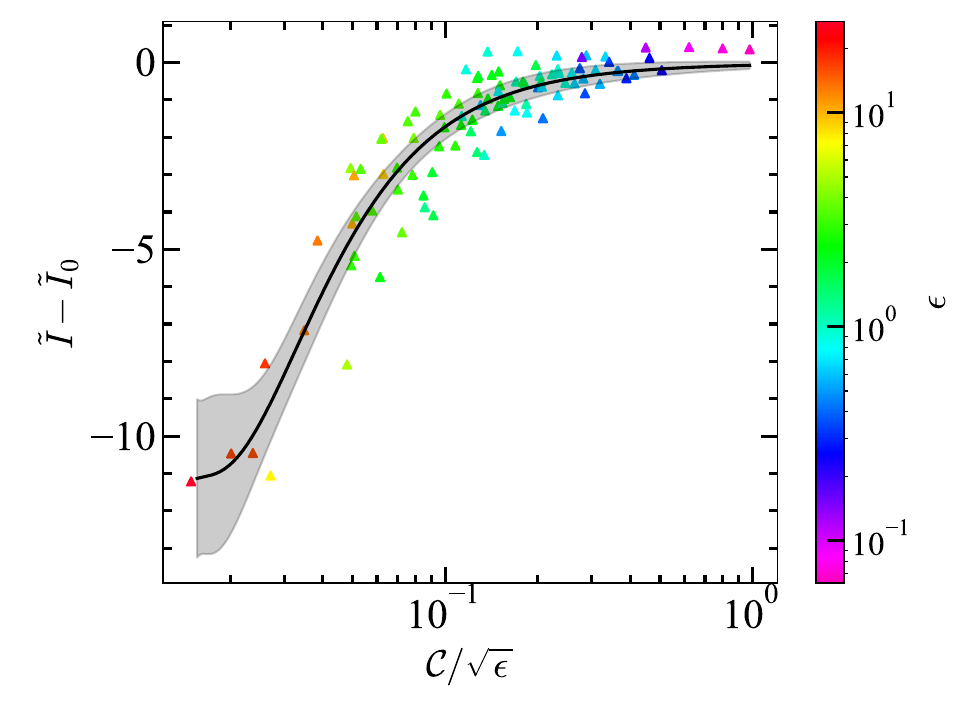}
    }
    \caption{Contribution on $\tilde{I}$ due to presence of magnetic fields are plotted as functions of $\mathcal{C}$ for differentially rotating systems with $A=8$.}
    \label{Fig: Field1+2}
\end{figure}

For systems with fixed $A$, the presence of a non-zero magnetic field modifies the stellar structure and systematically shifts the configurations toward lower $\mathcal{C}$ values. The amount of deviation depends on the initial $\mathcal{C}$ of the NS, as evident in Fig.~\ref{Fig: mag_field_effects} (see Sec.~\ref{sec: discussion} for more details). For a given magnetic field strength, configurations with lower $\mathcal{C}$ exhibit larger deviation from the non-magnetized configuration. To quantify the effect of magnetic fields, we decompose the total contribution into rotational and magnetic components as follows:
\begin{align}\label{Eq: MI magnetic}
    \tilde{I}(\mathcal{C},A,B_\mathrm{max}) = \tilde{I}_0(\mathcal{C},A) + \tilde{I}(\mathcal{C},B_\mathrm{max}).
\end{align}
Here, the first term on the right-hand side encodes the contribution arising from differential rotation and follows the established quasi-universal relation discussed in the previous section. Fig.~\ref{Fig: Field1+2}(a) illustrates $\tilde{I}(\mathcal{C},B_\mathrm{max})$, obtained by subtracting $\tilde{I}_0(\mathcal{C},A)$ from $\tilde{I}(\mathcal{C},A,B_\mathrm{max})$, as a function of $\mathcal{C}$ and $B_\mathrm{max}$ for a fixed value of $A$. The distribution of data points is relatively sparse in this representation, making it difficult to infer a robust analytic fitting form directly. To preserve the functional structure of the rigid-rotation fit as in Eq.~\eqref{Eq: rigid_fit}, the magnetic contribution is assumed to follow an analogous parametric form, with coefficients depend implicitly on the magnetic field strength. 

In this context, to improve the scaling behavior and obtain an underlying self-similarity, we introduce the dimensionless magnetic ellipticity $\epsilon = B_\mathrm{max}^2R^4/M^2$~\citep{Haskell:2007bh}. which characterizes the relative strength of magnetic deformation compared to gravitational force. Recasting the data in terms of the combined scaling variable $\mathcal{C}/\sqrt{\epsilon}$ significantly reduces scatter and reveals a clear monotonically increasing trend, as shown in Fig.~\ref{Fig: Field1+2}(b) This suggests that the magnetic contribution can be effectively parameterized as a function of $\mathcal{C}/\sqrt{\epsilon}$. Guided by this scaling behavior, we adopt the following polynomial form
\begin{equation}
    \tilde{I}(\mathcal{C},B_\mathrm{max}) = \sum_{i=1}^4 b_i \left(\frac{\mathcal{C}}{\sqrt\epsilon}\right)^{-i},
    \label{Eq: uni_fit}
\end{equation}
where $b_1-b_4$ are fitting coefficients that depend only on the rotational effect. This formulation has the additional advantage that it naturally enforces $\tilde{I}(\mathcal{C},B_\mathrm{max}) \to 0$ as $B_\mathrm{max} \to 0$; thereby recovering the purely rotational contribution and maintaining consistency with the non-magnetic limit discussed in the aforementioned sections. In Table~\ref{Tab: Fitted coeffictient B}, we list all the fitted values of $b_i$ for different $A$. 

\begin{table}[htbp]
    \centering
    \caption{Table of fitted magnetic coefficients for Eq.~\eqref{Eq: uni_fit} with different degrees of differential rotation.}
    \setlength{\tabcolsep}{3pt}
    \begin{tabular}{|l|c|c|c|c|c|}
        \hline
         & $b_1\, (10^{-2})$ & $b_2\, (10^{-2})$ & $b_3\, (10^{-4})$ & $b_4\, (10^{-6})$ & $\chi^2_{\rm red}$\\
        \hline
       $A\to\infty$ & $-8.8151$ & $-2.9040$ & $7.4394$ & $-4.9661$ & $0.0072$\\
        \hline
       $A=10$ & $-8.5265$ & $-1.4490$ & $3.9148$ & $-2.7267$ & $0.0008$\\
        \hline
       $A=8$ & $-5.9554$ & $-1.4277$ & $3.2389$ & $-2.0112$ & $0.0007$\\
        \hline
       $A=6$ & $-3.9484$ & $-1.1837$ & $2.0050$ & $-0.8221$ & $0.0005$\\
       \hline
    \end{tabular}
    \label{Tab: Fitted coeffictient B}
\end{table}

\begin{figure}[htbp]
    \centering
    \includegraphics[width=\columnwidth]{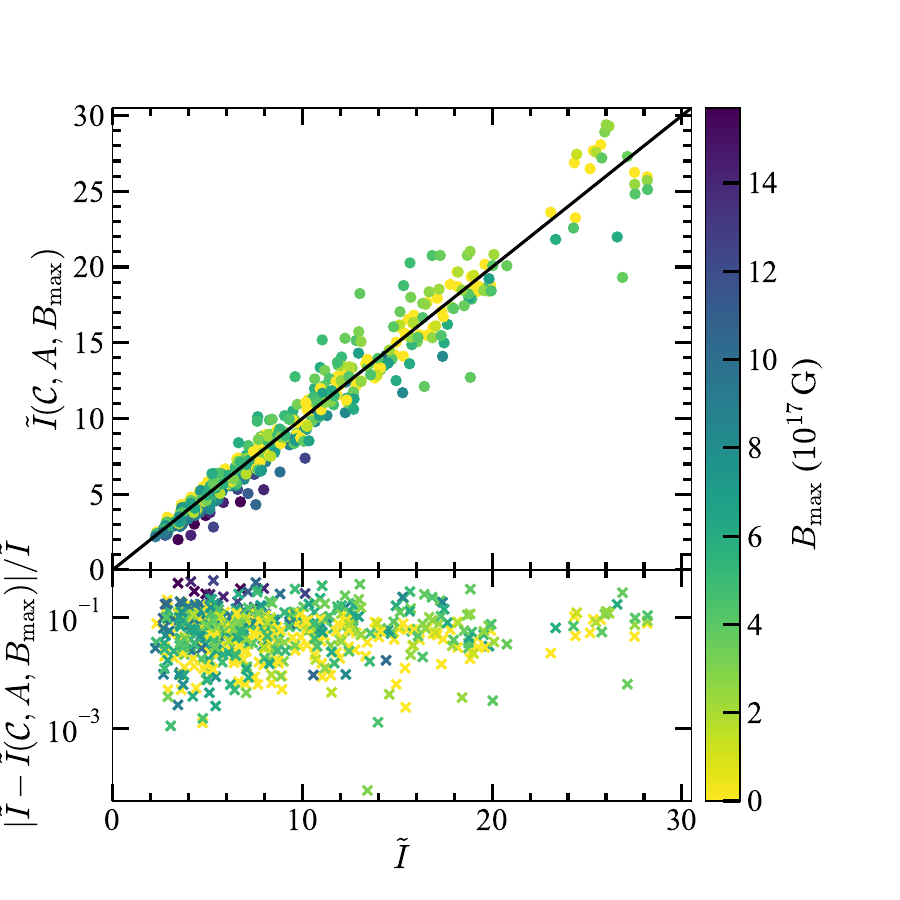}
    \caption{\textit{Top}: Comparison between $\tilde{I}$ and $\tilde{I}(C,A,B_{\rm max})$ obtained from the unified relation is shown. The unity line is used as a reference to visualize the agreement between the two quantities. \textit{Bottom}: The relative errors for all the configurations are plotted. The average (maximum) relative error is estimated to be $\sim0.08\ (0.5)$.} 
    \label{Fig: collapse}
\end{figure}

To establish the functional dependence of the coefficients on $A$, we again fit each coefficient $b_i$ as a function of $A$ following a similar analytical mapping shown in Eq.~\eqref{a_i_mapping}. Incorporating this dependence into the parameterization leads to the following generalized expressions
\begin{equation}\label{Eq: b coeff}
    \begin{aligned}
        b_1 &= -0.0872 - \frac{0.1829}{A} + \frac{2.6105}{A^2} + \frac{1.5044}{A^3}, \\
        b_2 &= -0.0286 - \frac{0.1485}{A} + \frac{4.4154}{A^2} - \frac{17.5769}{A^3}, \\
        b_3 &= 0.0007 + \frac{0.0039}{A} - \frac{0.1084}{A^2} + \frac{0.3969}{A^3}, \\
        b_4 &= -4.9 \times 10^{-6} - \frac{0.00003}{A} + \frac{0.0007}{A^2} - \frac{0.0023}{A^3},
    \end{aligned}
\end{equation}
with each case yields $\chi^2_{\rm red}<0.03$. To validate the fitted relations, we compute $\tilde{I}(C,A,B_{\rm max})$ from Eq.~\eqref{Eq: MI magnetic} for all magnetized and non-magnetized configurations and compare the resulting values with the corresponding original $\tilde{I}$. Fig.~\ref{Fig: collapse} shows a comparison between the predicted values $\tilde{I}(C,A,B_{\rm max})$ and the original data. It is evident that the fitted relations reproduce the original results with high accuracy with average (maximum) relative error $\sim 0.08\ (0.5)$.

\section{Astrophysical Implications}\label{sec: astro_imp}
In this section, we show two simple astrophysical applications in magnetars and GWs of the unified relation mentioned in Eq.~\eqref{Eq: MI magnetic}. 

\subsection{Magnetars}
We consider a sample of 25 magnetars from the McGill Magnetar Catalog~\citep{Olausen} and examine the impact of magnetic-field-induced corrections to $I$ on the inferred spin-down luminosity ($\dot{E}$), given by
\begin{equation}\label{spin-down-lim}
    \dot{E} = 4\pi^2I\frac{\dot{P}}{P},
\end{equation}
where $P$ and $\dot{P}$ denote the rotation period and its time derivative, respectively. As a reference model, we assume a uniformly rotating magnetar with mass $M = 1.4\,\rm M_\odot$ and radius $R = 10$\,km; thereby obtain $I \approx 10^{45}\rm\,g\,cm^2$, consistent with the canonical value commonly adopted in magnetar studies, which was originally used to compute $\dot{E}$ tabulated in the catalog. Considering it as the baseline, we then replace this value in Eq.~\eqref{spin-down-lim} with the magnetic-field-corrected moment of inertia, $I_{\rm corr} =\tilde{I}(\mathcal{C},A,B_{\max})\times M^3$, using the unified relation of Eq.~\eqref{Eq: MI magnetic}. This allows us to quantify how structural modifications due to magnetic fields affect $\dot{E}$.

As inference on the surface magnetic fields ($B_\mathrm{s}$) cannot constrain $B_{\max}$, we relate them through the phenomenological scaling $B_{\rm max}=\eta B_\mathrm{s}$, where $\eta$ is a dimensionless tuning parameter. The values of $\eta$ are chosen in such a way that the resulting $B_{\max}$ remains within, or close to, the range over which the unified relation has been calibrated; thereby minimizing extrapolation uncertainties.

\begin{figure}[htbp]
    \centering
    \includegraphics[width=\columnwidth]{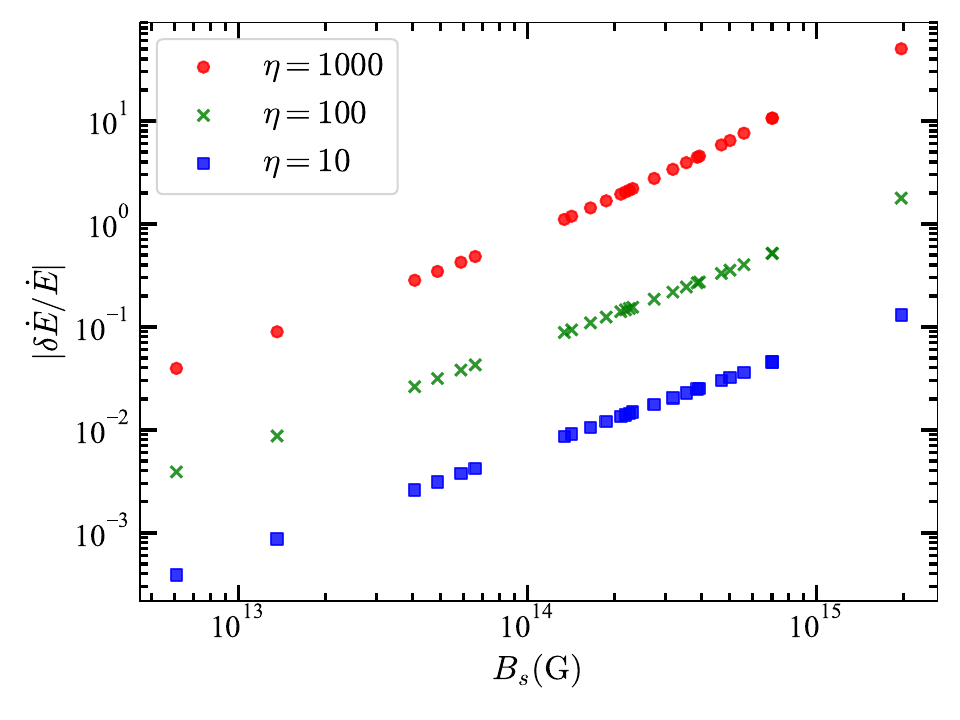}
    \caption{Fractional change in $\dot{E}$ of magnetars plotted as a function of $B_\mathrm{s}$ for constant $J$ configurations with $M=1.4\rm\,M_\odot$ and $R=10$\,km. For each source, the interior magnetic field increases with the tuning parameter $\eta$.} 
    \label{Fig: magnetar}
\end{figure}
Fig.~\ref{Fig: magnetar} shows the fractional variation in the spin-down luminosity $\delta\dot{E}/\dot{E}$, as a function of $B_\mathrm{s}$ for all sources in the dataset. It is evident that it can change in many orders-of-magnitude depending on $\eta$. This result demonstrates that magnetic field contributions can significantly influence the parameters derived from astrophysical quantities, even when the observable timing parameters $(P,\dot{P})$ remain unchanged. In this proof-of-concept analysis, replacing $I$ with $I_{\rm corr}$ effectively corresponds to comparing equilibrium stellar configurations along a constant $J$ sequence. Thus, neglecting such corrections may introduce systematic biases in the interpretation of highly magnetized NSs.

\subsection{Gravitational waves}
The unified relation can also be applied to the post-merger GW emission from binary NS merger events like GW170817. In this context, we consider a canonical remnant NS with $M=2.7\rm\,M_\odot$ and $R=12$\,km~\citep{Margalit:2017dij,gw170817}. We then identify the combinations of $A$ and $B_{\rm max}$ that are consistent with the effective remnant spin frequency ($f$) in the range $1-2$\,kHz.\footnote{This frequency interval is inspired from the fact that the dominant quadrupolar emission frequency occurs at $\sim2f$, corresponding to frequencies of $2-4$\,kHz~\citep{Bauswein:2015vxa}. Note that, this approximation is primarily applicable to rigidly rotating remnants and introduces an additional systematic uncertainty in the presence of differential rotation, which we do not attempt to quantify here.} For the allowed region in ($A,B_{\rm max}$) parameter space, we further estimate the corresponding rotational kinetic energy ($T_{\rm rot}$).

\begin{figure}[htbp]
    \centering
    \includegraphics[width=\columnwidth]{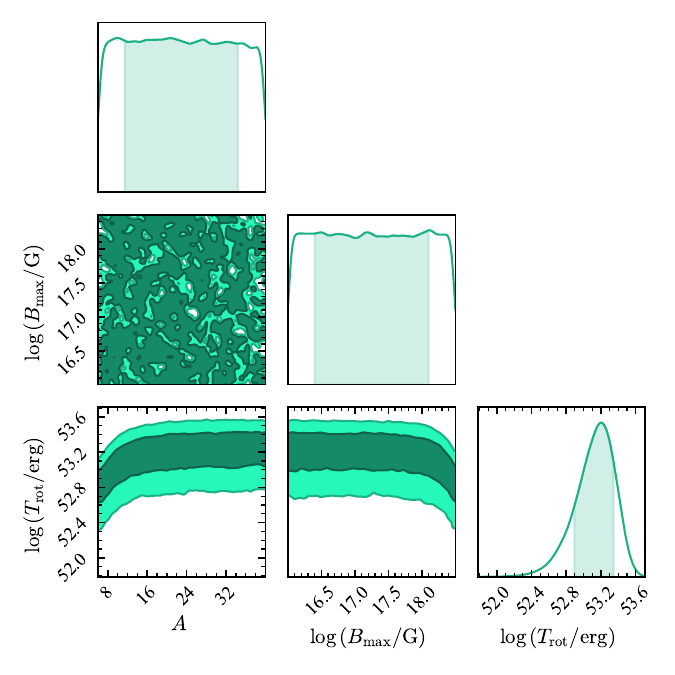}
    \caption{Corner plot showing the correlation among $T_{\rm rot}$, $A$, and $B_{\max}$. The individual posteriors of $A$ and $B_{\max}$ are essentially flat except for strong differential rotation or high magnetic field regime.}
    \label{Fig: MCMC}
\end{figure}

Fig.~\ref{Fig: MCMC} shows a corner plot to show the the correlation among $T_{\rm rot}$, $A$, and $B_{\max}$, along with their posterior distributions. We find that, the assumed frequency range results in a degenerate band in ($A,B_{\rm max}$) plane, indicating that strong differential rotation or intense magnetic fields can reduce $T_{\rm rot}$ through their influence on $I(\mathcal{C},A,B_{\rm max})$. The resulting posterior distribution of $T_{\rm rot}$ is well constrained around $10^{53}$\,ergs, in agreement with the estimate of~\cite{Margalit:2017dij}. However, the marginal posterior distributions of $A$ and $B_{\rm max}$ remain largely flat for most of the parameter space; thereby indicating that constraining $f$ alone is insufficient to break the degeneracy between $A$ and $B_{\rm max}$; highlighting the need for additional observables such as those accessible through post-merger GW spectroscopy.


\section{Discussion}\label{sec: discussion}

In this work, we have investigated the combined effects of differential rotation and toroidal magnetic fields on the macroscopic properties of NSs, focusing on two fundamental observables: $I$ and $\mathcal{C}$. With the help of \textsc{xns} code utilizing 10 different EoSs, we have obtained equilibrium configurations of NSs. For sequences of constant $J$, we have demonstrated that the relation
\begin{align}
    \tilde{I}(\mathcal{C},A,B_\mathrm{max}) = \sum_{i=1}^4 \left[a_i(A) + b_i(A)\left(\frac{B_\mathrm{max}R^2}{\sqrt{G}M}\right)^i\right]\mathcal{C}^{-i},
\end{align}
remains valid not only for uniformly rotating configurations, as previously shown by \cite{Breu:2016ufb}, but also for highly magnetized differentially rotating NSs. Unlike the uniformly rotating case, the coefficients $a_i$ and $b_i$, given by the set of Eqs.~\eqref{Eq: a coeff} and~\eqref{Eq: b coeff}, explicitly depend on the degree of differential rotation, naturally recovering the uniform rotation limit as $A\to\infty$. Finally, we have demonstrated the astrophysical relevance of this generalized relation by quantifying the impact of differential rotation and magnetic fields on two representative applications: luminosities of magnetars and NS merger event GW170817.

In case of differential rotation, we find that systems with fixed $\mathcal{C}$ tend to shift towards lower $\tilde{I}$ values. The magnitude of this shift depends on $A$. This trend arises from the redistribution of $J$ induced by differential rotation. As $A$ decreases, $\Omega$ profile becomes increasingly concentrated toward the NS core, enabling the central regions to provide a larger fraction of the centrifugal support. Consequently, NSs with the same $\mathcal{C}$ can be sustained with a less extended outer envelope than in the uniformly rotating case. Since to the leading order, $J\sim I\Omega_c$, an increase in $\Omega_c$ corresponds to a reduction in $I$ for fixed $J$. For the range of $A$ considered in this work, the change in $I$, dominates over the change in $M$, leading to an overall reduction in $\tilde{I}$. The deviation from the uniform rotation relation increases with with increasing differential rotation.

To understand the effects of magnetic fields, we focus on a single configuration with fixed $J$ and central density $\rho_\mathrm{c}$. Magnetic fields provide extra pressure through Lorentz force to support the NS; thereby reducing the amount of $M$ that can be sustained in equilibrium for a given $\rho_{\mathrm{c}}$. At the same time, the enhanced pressure supports an increase in the $R$. As a result, $\mathcal{C}=M/R$ decreases and the NS become less compact. Simultaneously, magnetic field modifies the rotation profile via $J$ redistribution, generally increasing $\Omega_c$ relative to the outer layers. As $\tilde{I}=I/M^3$, it can be expressed as $\tilde{I}=J/\Omega_\mathrm{c}M^3$. Therefore, the evolution of $\tilde{I}$ is governed by the competing effects of the changes in $M$ and $\Omega_{\mathrm{c}}$. Depending on their relative magnitudes, $\tilde{I}$ deviates from the non-magnetized configuration as illustrated in Fig.~\ref{Fig: mag_field_effects}.

Additionally, we compute deformability ($e$) for our rotating configurations, defined as $e=c_B|B|^2$, where $c_B$ is termed as the distortion coefficient~\citep{Soldateschi:2021cxq}.\footnote{Note that, the dimensional ellipticity $\epsilon$, introduced in Sec.~\ref{mag_fields}, is a standard variant of $e$ with the scaling $R^4/M^2$~\citep{Haskell:2007bh}.} For static systems with predominantly toroidal magnetic fields, \cite{Soldateschi:2021cxq} derived a quasi-universal relation for $c_B$ of the form
\begin{equation}
    c_{\rm B} = 0.25R_{10}^{5.03}M_{\rm 1.6}^{-2.07},
\label{solda_fit}
\end{equation}
where $R_{10}=R/10{\rm\, km}$ and $M_{1.6}=M/1.6\rm\,M_\odot$. Motivated by this result, we examine whether an analogous scaling describes our uniformly and differentially rotating $(A=6)$ equilibrium sequences. Fitting the same functional dependence to our models yields
\begin{equation}
c_{\rm B} =
\left\{
\begin{aligned}
&(0.172\pm0.023) R_{10}^{(3.17\pm0.39)}M_{\rm 1.6}^{(-2.18\pm0.17)};\ A\to\infty\\
&(0.167\pm0.020) R_{10}^{(3.31\pm0.40)}M_{\rm 1.6}^{(-2.25\pm0.17)};\ A=6
\end{aligned}
\right.
\label{our_fit}
\end{equation}
where the quoted uncertainties correspond to the 1$\sigma$ confidence interval. We find that Eqs.~\eqref{solda_fit} and \eqref{our_fit} demonstrate similar mass dependencies (within the uncertainties) although differing in radial exponents. This comparison should, however, be interpreted with caution. The relation of \cite{Soldateschi:2021cxq} was calibrated exclusively for non-rotating configurations, whereas our fit is obtained from uniformly and differentially rotating magnetized models. The difference in the radial scaling is therefore likely to reflect rotational modifications to the stellar structure. Eq.~\eqref{our_fit} provides only a quantitative comparison to the existing result, rather than a proposed quasi-universality on $c_{\rm B}$ which requires a thorough systematic study on diverse range of rotational and stellar parameters. Such investigation is beyond the scope of the present study and is left for a future work.

Finally, we highlight some limitations of our work. First, we assume purely toroidal fields and our treatment of magnetic corrections relies on polynomial fits in $\mathcal{C}$. Although, this preserves the original structure of universal relation identified by \cite{Breu:2016ufb}, it remains as a phenomenological ansatz and might require additional parametrizations to isolate the effects of $A$ and $B_{\rm max}$ in astrophysical applications. Second, the unified relation is trained upon equilibrium configurations spanning a broad range of $j$, although residing on a constant $J$ sequence. This selection criteria is necessary to isolate the effect of rotation profile and accurately quantify the effects of differential rotation (see \citealt{Taylor:2019hle} for a detailed discussion). Therefore, appropriate caution is needed while applying this relation for astrophysical scenarios in which $J$ is not conserved, such as systems undergoing magnetic braking or angular momentum losses driven by GW emission or neutrino outflows. Lastly, our equilibrium models are computed for zero temperature, neglecting any thermal support that could be significant in realistic post-merger phase before the remnant cools down.

Overall, the proposed framework represents a significant step toward constructing quasi-universal relations for more realistic proto-NS models and provides a foundation for future extensions incorporating additional physical effects.

\begin{acknowledgements}
The authors thank the anonymous reviewer for their insightful comments, which have helped to improve the quality of this manuscript, particularly for the suggestions to compare our results with previous studies. P.M. thanks Alejandro Casallas Lagos of University of Warsaw for useful discussion on GW merger event. S.K. acknowledges funding from the National Science Centre, Poland (grant no. 2023/49/B/ST9/02777).
\end{acknowledgements}

\section*{Data availability}
The data underlying this article were generated with the help of {\sc xns} code; whereas magnetar data is obtained from publicly available catalog. The derived data generated in this research will be shared on reasonable request to the corresponding author.

\bibliographystyle{aa}
\bibliography{reference}

\end{document}